\documentclass{elsart}

\usepackage{graphicx}
\usepackage{amsmath,amssymb}
\usepackage{times}
\usepackage{bm}

\newcommand{\eq}[1]{\begin{equation}#1\end{equation}}

\newcommand{\ket}[1]{\ensuremath{\,|{#1}\rangle}}

\newcommand{\op}[1]{\ensuremath{\mathrm{#1}}}
\newcommand{\adj}[1]{\ensuremath{{{#1}}^{\dag}}}
\newcommand{\corr}[1]{\ensuremath{\widetilde{#1}}}

\newcommand{\ii}{\ensuremath{\mathrm{i}}}
\newcommand{\dd}{\ensuremath{\mathrm{d}}}

\renewcommand{\vec}[1]{\ensuremath{\bm{#1}}}

\newcommand{\gO}{\ensuremath{\op{g}}}

\newcommand{\qO}{\ensuremath{\op{q}}}
\newcommand{\rO}{\ensuremath{\op{r}}}

\newcommand{\CO}{\ensuremath{\op{C}}}
\newcommand{\CCO}{\ensuremath{\adj{\op{C}}}}

\newcommand{\HO}{\ensuremath{\op{H}}}

\newcommand{\OO}{\ensuremath{\op{O}}}

\newcommand{\TO}{\ensuremath{\op{T}}}
\newcommand{\VO}{\ensuremath{\op{V}}}

\newcommand{\qOV}{\ensuremath{\vec{\op{q}}}}
\newcommand{\rOV}{\ensuremath{\vec{\op{r}}}}

\newcommand{\sigmaOV}{\ensuremath{\vec{\op{\sigma}}}}

\newcommand{\tensorO}{\ensuremath{\op{s}_{12}}}

\newcommand{\UCOM}{\ensuremath{\textrm{UCOM}}}
\newcommand{\intr}{\ensuremath{\textrm{int}}}

\newcommand{\cm}{\ensuremath{\textrm{cm}}}
\newcommand{\elem}[2]{\ensuremath{{}^{#2}\text{#1}}}

\begin{document}

\begin{frontmatter}

\title{Nuclear Structure in the UCOM Framework:\\ From Realistic Interactions to Collective Excitations\thanksref{sfb}}

\thanks[sfb]{This  work is supported by the Deutsche Forschungsgemeinschaft (DFG) through contract SFB 634.}

\author{R. Roth, H. Hergert, N. Paar, P. Papakonstantinou}

\address{Institut f\"ur Kernphysik, Technische Universit\"at Darmstadt,\\ Schlossgartenstr. 9, 64289 Darmstadt, Germany}

\begin{abstract}
The Unitary Correlation Operator Method (UCOM) provides a means for nuclear structure calculations starting from realistic NN potentials. The dominant short-range central and tensor correlations are described explicitly by a unitary transformation. The application of UCOM in the context of the no-core shell model provides insight into the interplay between dominant short-range and residual long-range correlations in the nuclear many-body problem. The use of the correlated interaction within Hartree-Fock, many-body perturbation theory, and Random Phase Approximation gives access to various nuclear structure observables throughout the nuclear chart.
\end{abstract}


\end{frontmatter}

\section{Introduction}

In recent years several realistic nucleon-nucleon interactions like the Argonne V18 \cite{WiSt95} and the CD Bonn potentials as well as interactions derived from a chiral effective field theory \cite{EnMa03,EpNo02} have been constructed on the basis of high-precision nucleon-nucleon scattering data. These potentials are used in \emph{ab initio} nuclear structure calculations throughout the p-shell, e.g., in the framework of the Green's function Monte Carlo method or the no-core shell model \cite{BaMi03}. The use of these realistic potentials for nuclear structure studies in heavier nuclei poses an enormous challenge. Traditional many-body methods, like the Hartree-Fock approach or the Random Phase Approximation, cannot be used in connection with the bare NN interaction. The reason is the inability of the simple model spaces underlying these approaches to describe the dominant short-range correlations, which are present in the exact many-body eigenstates.

The two most important types of many-body correlations are those induced by the short-range repulsion and the strong tensor part of the NN interaction. Already in the deuteron their presence is evident: (\emph{i}) The probability amplitude for finding the two nucleons at very small distances is strongly depleted as a result of the short-range repulsive interaction. (\emph{ii}) Apart from the $L=0$ component the ground state exhibits a $L=2$ admixture which is essential for the binding and is generated by the tensor part of the NN interaction. Also for heavier nuclei, these central and tensor correlations have a dominant impact on the structure of the many-body state. Neither of these correlations can be described properly by a single or a superposition of few Slater determinants. Therefore, a naive inclusion of the bare realistic NN-potential into a Hartree-Fock-type calculation has to fail. 

For nuclei beyond the p-shell one is bound to use simplified model spaces for an approximate solution of the many-body problem. Therefore, the short-range correlations have to be accounted for explicitly, e.g., by transforming the bare realistic interaction into an effective interaction adapted to the available model space. One possible approach to construct a phase-shift equivalent effective interaction is the Unitary Correlation Operator Method discussed in Section \ref{sec:ucom}. The resulting correlated interaction $\VO_{\UCOM}$ can then be used as a universal input for different many-body approaches, ranging from the no-core shell model (Section \ref{sec:ncsm}) over Hartree-Fock and many-body perturbation theory (Section \ref{sec:hf}) to the Random Phase Approximation (Section \ref{sec:rpa}).

\section{Unitary Correlation Operator Method}
\label{sec:ucom}

The basic idea of the Unitary Correlation Operator Method (UCOM) is to include the dominant correlations into the many-body state by means of a unitary transformation \cite{FeNe98,NeFe03,RoNe04,RoHe05}. Starting from an uncorrelated many-body state $\ket{\Psi}$, in the simplest case just a Slater determinant, a correlated state $\ket{\corr{\Psi}}$ is defined through the application of the unitary correlation operator $\CO$:
\eq{
  \ket{\corr{\Psi}} = \CO \ket{\Psi} \;.
}
Alternatively, one can perform a similarity transformation of the operators of all relevant observables (e.g. the Hamiltonian, coordinate and momentum space densities, transition operators, etc.):
\eq{
  \corr{\OO} = \CCO \OO \CO \;.
}
Due to unitarity both approaches are equivalent. For most many-body calculations the formulation through correlated operators is, however, more convenient.

The correlation operator $\CO$ is decomposed into a central correlator $\CO_r$ and a tensor correlator $\CO_{\Omega}$, reflecting the two dominant types of correlations in the nuclear many-body problem:
\eq{
  \CO = \CO_{\Omega} \CO_{r} 
  = \exp\!\!\Big[-\ii \sum_{i<j} \gO_{\Omega}(ij) \Big] 
    \exp\!\!\Big[-\ii \sum_{i<j} \gO_{r}(ij) \Big]\;.
}
Both operators are defined as exponentials of Hermitian two-body generators $\gO_{\Omega}$ and $\gO_{r}$, respectively. They are given in a closed analytic form which reflects the mechanism by which correlations are induced by the interaction.

The task of the central correlator $\CO_r$ is to generate the hole in the two-body density distribution at small particle distances caused by the repulsive core in the central part of the interaction. Pictorially speaking, $\CO_r$ has to shift pairs of particles that are closer than the core radius apart from one another. The two-body generator for this distance-dependent shift can be written as $\gO_r = \frac{1}{2} [s(\rO) \qO_r + \qO_r s(\rO)]$, where $\qO_r = \frac{1}{2} [\qOV\cdot(\rOV/\rO) + (\rOV/\rO)\cdot\qOV]$ is the radial component of the relative momentum $\qOV$ of a particle pair. The function $s(r)$ determines the distance-dependence of the shift. It is large for small $r$ and vanishes at large distances.

The tensor correlation operator $\CO_{\Omega}$ has to generate the complex entanglement between the angular structure of the relative two-body states and the spin orientation. An essential ingredient is the component of the relative momentum $\qOV$ perpendicular to $\rOV$, the so-called orbital momentum $\qOV_{\Omega} = \qOV -
\tfrac{\rOV}{\rO}\;\qO_r$. The generator has the form $\gO_{\Omega}  =
\frac{3}{2}\vartheta(\rO) \big[(\sigmaOV_1\!\cdot\qOV_{\Omega})(\sigmaOV_2\!\cdot\rOV) + (\rOV \leftrightarrow \qOV_{\Omega})\big]$ which is similar to the tensor operator $\tensorO$. The function $\vartheta(r)$ describes the magnitude of the shift as a function of distance. 

For the following many-body calculations, the notion of correlated operators is advantageous. The operators of all observables under consideration have to be transformed consistently. Since the correlation operators are defined as exponentials of two-body operators, the correlated operators contain irreducible contributions for all particle numbers. We organize the different irreducible terms according to their rank in a cluster expansion
\eq{
  \corr{\HO} 
  = \CCO \HO \CO 
  = \corr{\HO}^{[1]} + \corr{\HO}^{[2]} + \corr{\HO}^{[3]} + \cdots \;.
}
Here we used the Hamiltonian $\HO=\TO+\VO$ as an example, but the same holds true for any other operator. If the range of the correlators is sufficiently small compared to the average particle distance in the many-body system, three-body and higher order terms in the cluster expansion are small and we can restrict ourselves to the two-body approximation
\eq{ \label{eq:Hcorr_twobody}
  \corr{\HO}^{C2} 
  = \corr{\TO}^{[1]} + \corr{\TO}^{[2]} + \corr{\VO}^{[2]} 
  = \TO + \VO_{\UCOM} \;,
}
where $\corr{\TO}^{[1]}=\TO$ and $\corr{\TO}^{[2]}$ are the one- and two-body contributions of the correlated kinetic energy, resp., and $\corr{\VO}^{[2]}$ is the two-body part of the correlated NN-potential. All two-body contributions are subsumed in the correlated interaction $\VO_{\UCOM}$. It is by construction \emph{phase-shift equivalent} to the original, uncorrelated NN-potential as long as the correlators have finite range. 

The remaining task is the determinantion of the correlation functions $s(r)$ and $\vartheta(r)$ entering into the generators of the unitary transformations. For each spin-isospin channel their parameters can be obtained from an energy minimization in the two-body system. This procedure and the optimal correlators for the Argonne V18 (AV18) potential are discussed in Ref. \cite{RoHe05}. The tensor correlation functions require a special treatment. Since it originates from the one-pion exchange, the tensor force is long-ranged, and so are the tensor correlations induced in the two-body system. In a many-body system, the long-range component of the tensor correlations between two nucleons is screened due to the presence of other nucleons. In anticipation of this effect, we restrict the range of the tensor correlator by a constraint on the integral of the correlation function, $I_{\vartheta} = \int \dd r\, r^2 \vartheta(r)$. Hence, only short-range correlations are described explicitly by the unitary transformation. Long-range correlations have to be covered by the many-body states---this will be illustrated in the following sections.

\section{No-Core Shell Model Calculations}
\label{sec:ncsm}

As a first application of the correlated realistic interaction $\VO_{\UCOM}$ we consider a straightforward no-core shell model diagonalization within a harmonic oscillator basis. The shell model basis itself is able to describe part of the many-body correlations, depending on the size of the model space. Hence the dependence of the energy on the model-space size provides information on the role of short-range correlations and on the contribution from residual long-range correlations. For the calculations we employ the translationally invariant no-core shell model code developed by Petr Navr\'atil \cite{NaKa00}, but without using the Lee-Suzuki transformation. The computation of the relevant two-body matrix elements of $\VO_{\UCOM}$ in the harmonic oscillator basis and further results are discussed in Ref. \cite{RoHe05}.
  
\begin{figure}[t!]
  \begin{center}
  \includegraphics[width=0.4\textwidth]{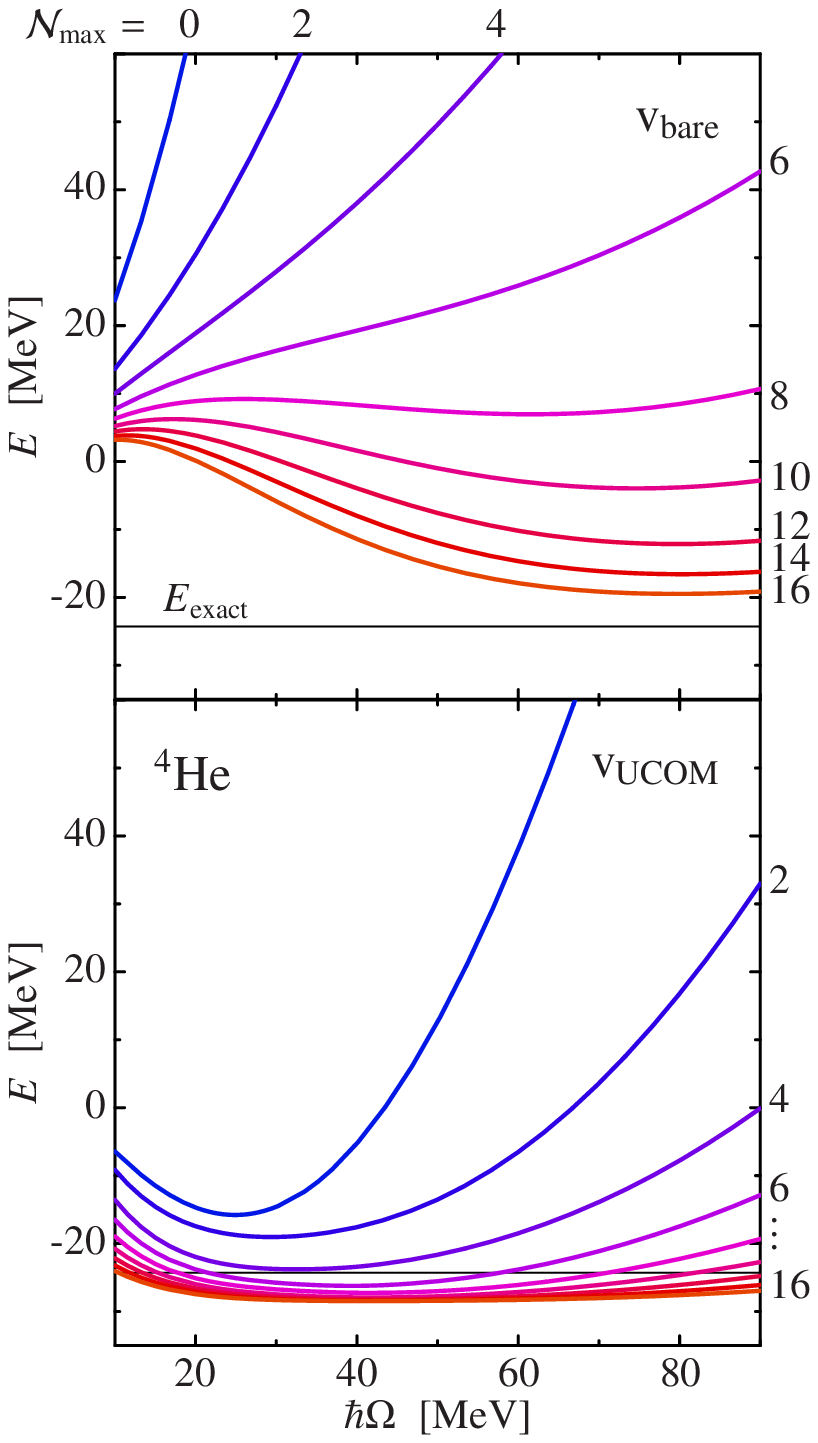}
  \qquad
  \raisebox{7ex}{\includegraphics[width=0.47\textwidth]{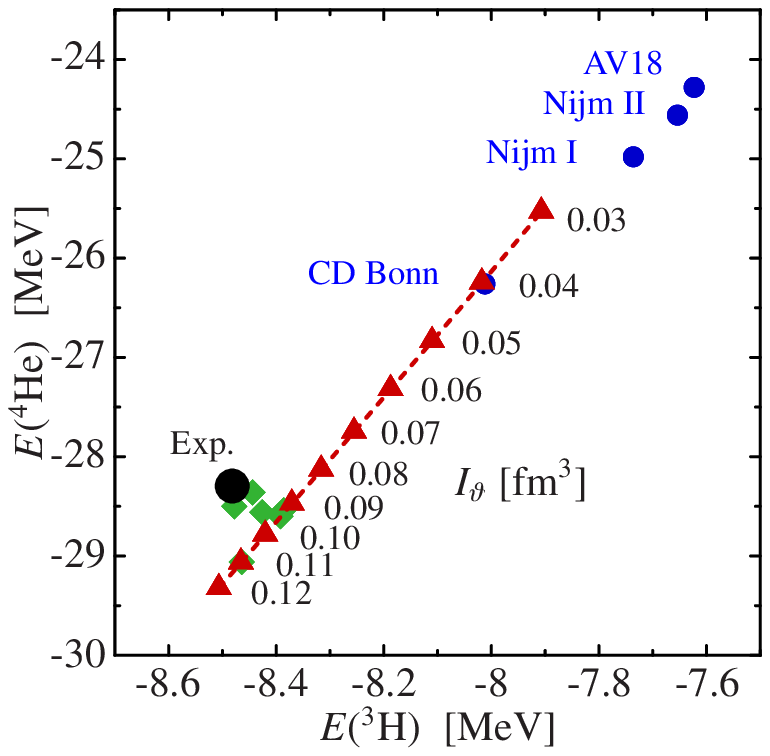}}
  \end{center}
  \caption{Results of no-core shell model calculations using the correlated AV18 potential. Left panel: convergence of the ground state energy of \elem{He}{4} for bare (upper plot) and correlated AV18 potential (lower plot). Right panel: Tjon-line and dependence of the energy on the correlator range as described in the text (taken from \cite{RoHe05}).}
  \label{fig:ncsm}
\end{figure}

Figure \ref{fig:ncsm} shows the ground state energy of \elem{He}{4} as a function of the oscillator parameter $\hbar\Omega$ for different sizes of the model space, characterized by the maximum relative oscillator quantum number $\mathcal{N}_{\max}$. The upper panel corresponds to a calculation with the bare AV18 potential. Evidently, even for the largest feasible model spaces, the energy is not yet converged. The reason is that a full description of short-range central and tensor correlations requires even larger model spaces, which are computationally not tractable. The picture changes if we use $\VO_{\UCOM}$, i.e., include the unitary transformation of the Hamiltonian. The convergence is dramatically improved since the short-range central and tensor correlations are now treated explicitly by the unitary correlation operator. Note that a bound nucleus is already obtained with a single Slater determinant (i.e. $\mathcal{N}_{\max}=0$). With increasing size of the model space, the ground state energy is lowered further. This is the result of the improved description of long-range correlations---not accounted for by the unitary transformation---by the model space.
  
A second interesting aspect is illustrated on the right-hand side of Fig. \ref{fig:ncsm}, where the converged ground state energies of \elem{H}{3} and \elem{He}{4} are plotted. Each data point corresponds to a different interaction. The exact energies for the different bare NN-interactions, like the AV18, the CD Bonn and the Nijmegen interactions (circles), fall onto the so-called Tjon-line \cite{NoKa00} but are far away from the experimental point. Three-nucleon interactions (diamonds) are needed to obtain binding energies in the experimental region. The exact energies for the correlated interaction $\VO_{\UCOM}$ based on AV18 (triangles) depend on the range $I_{\vartheta}$ of the triplet-even tensor correlation function. With increasing range the energy is lowered and the full Tjon-line is mapped out. This is related to the omission of three-body (and higher-order) terms in the cluster expansion of the correlated Hamiltonian. If these terms were included, the energies would be exactly the same, independent of the correlator range, because of the unitarity of the transformation. The fact that the range of the tensor correlator can be chosen such that the energies are close to experiment (e.g. for $I_{\vartheta}=0.09\,\text{fm}^3$) can be explained by a cancellation between genuine three-body forces and the induced three-body contributions of the cluster expansion. In other words, the impact of the net three-body force on the binding energies can be minimized by a proper choice of the correlator range.

\section{Hartree-Fock \& Many-Body Perturbation Theory}
\label{sec:hf}

Using the $\VO_{\UCOM}$ interaction fixed within the no-core shell model we perform Hartree-Fock (HF) calculations of nuclear ground states throughout the nuclear chart. Since the HF many-body state (Slater determinant) alone is not able to describe any many-body correlations, the use of bare realistic interactions does not lead to bound nuclei. The explicit inclusion of the short-range correlations, e.g., via the unitary correlation operators is inevitable. 

\begin{figure}
  \begin{center}
  \includegraphics[width=0.8\textwidth]{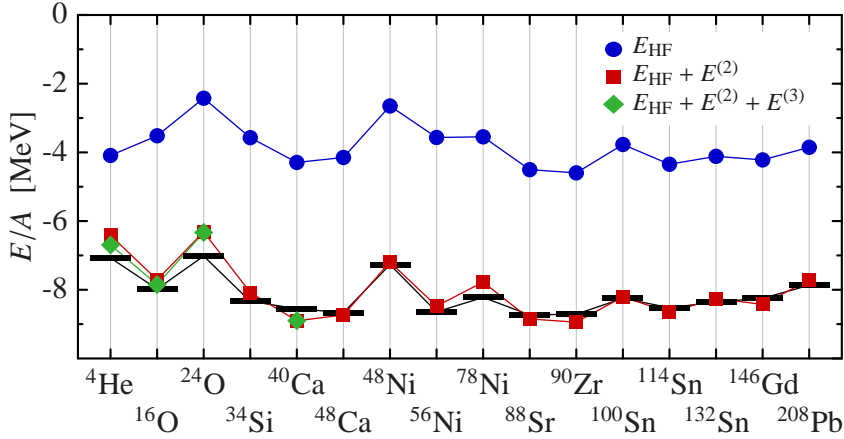}
  \end{center}
  \caption{Ground state energy of various closed shell nuclei obtained with the correlated AV18 potential within a HF calculation (circles) and in HF + MBPT (squares and diamonds) in comparison to experimental binding energies (bars) (taken from \cite{RoPa05}).}
  \label{fig:hf+mbpt}
\end{figure}

We have implemented the HF scheme in the harmonic-oscillator representation, using the translationally invariant Hamiltonian $\HO_{\intr}=\TO-\TO_{\cm}+\VO_{\UCOM}$, where $\VO_{\UCOM}$ contains charge-dependent and Coulomb terms \cite{RoPa05}. The results for ground state energies of closed-shell nuclei ranging from \elem{He}{4} to \elem{Pb}{208} are depicted in Fig. \ref{fig:hf+mbpt}. The optimal correlator for $I_{\vartheta}=0.09\,\text{fm}^3$ is used, and the single-particle basis includes 13 major oscillator shells. Evidently, the HF binding energies are significantly smaller than the experimental ones. This is not surprising, since residual long-range correlations as they appeared in the no-core shell model calculations cannot be described by the HF ground state.

An estimate for the impact of residual long-range correlations on the binding energies can be obtained within many-body perturbation theory. The evaluation of the second and third order perturbative contributions on top of the HF result is straightforward \cite{RoPa05}. Figure \ref{fig:hf+mbpt} summarizes the results for the ground state energies including second order correlations (for light nuclei also third order). Again, $13$ major oscillator shells are included to obtain a satisfactory degree of convergence for the perturbative correction. The agreement with the experimental binding energies per nucleon is remarkably good throughout the whole mass range. The absence of any systematic deviation for larger mass numbers proves that the cancellation between genuine three-body force and induced three-body contributions, which we observed in the no-core shell model for light isotopes, works throughout the nuclear chart. Furthermore, the calculations establish the perturbative character of the long-range correlations. Note that a perturbative treatment of the short-range correlations is not possible---in our approach they are covered by the unitary correlation operators from the outset.

However, the good agreement with experimental data does not hold for all observables. The charge radii obtained in HF for heavier nuclei are too small in comparison to experiment \cite{RoPa05}. The inclusion of perturbative corrections improves the result but still leaves deviations of up to $1\,\text{fm}$ for the heaviest nuclei. This is an indication that a net three-body force is needed to reproduce all observables, although its impact on the energy might be small. This issue is the topic of future investigations.

\section{Random Phase Approximation}
\label{sec:rpa}

In addition to global ground state properties, collective excitations provide a valueable probe to understand the role of correlations in the nuclear many-body problem. We use the standard Random Phase Approximation (RPA) \cite{PaPa06} and its extensions \cite{PaRo06} to study the behavior of collective excitations based on $\VO_{\UCOM}$. Starting from the HF solution for the ground state we solve the RPA equations in a fully self-consistent way using the same intrinsic Hamiltonian as for the HF treatment. In this way the spurious center-of-mass mode appears fully decoupled at very low excitation energies of the order of $10$keV and the energy-weighted sum rules are fulfilled with maximum deviations of $\pm3\%$.  

\begin{figure}
  \begin{center}
  \includegraphics[width=\textwidth]{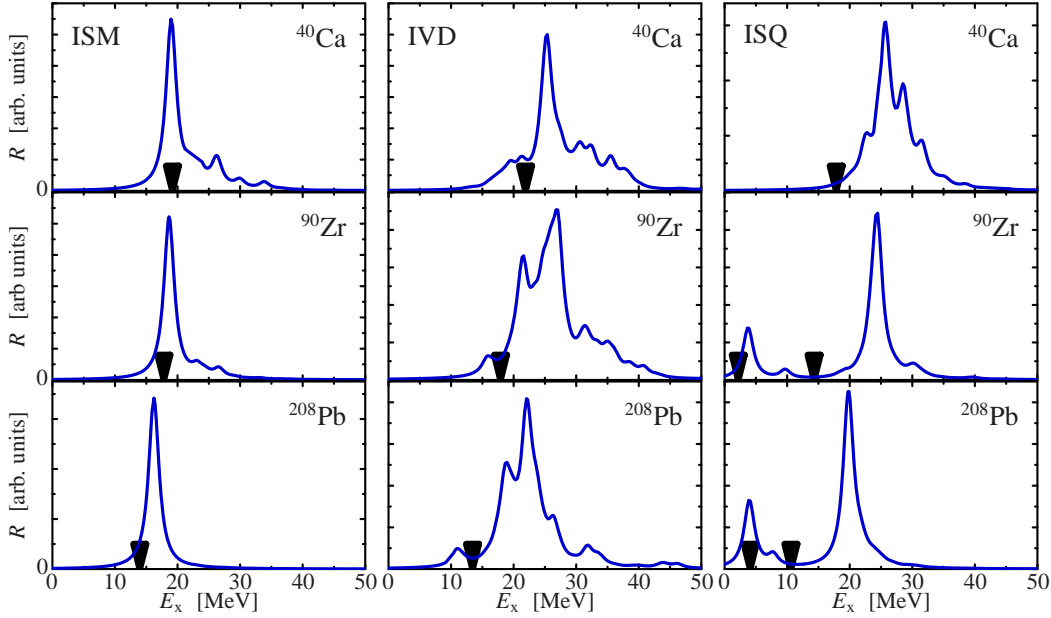}
  \end{center}
  \caption{RPA strengths distributions for isoscalar monopole (ISM), isovector dipole (IVD), and isoscalar quadrupole (ISQ) excitations in \elem{Ca}{40}, \elem{Zr}{90}, and \elem{Pb}{208} using $\VO_{\UCOM}$. The curves result from a Lorentzian folding of the discrete strength distributions and the black triangles indicate the centroid energies extracted from experiment (see also \cite{PaPa06}).}
  \label{fig:rpa}
\end{figure}

The results obtained for isoscalar monopole, isovector dipole, and isoscalar quadru\-pole excitations in  \elem{Ca}{40}, \elem{Zr}{90}, and \elem{Pb}{208} using the standard $\VO_{\UCOM}$ interaction are summarized in Figure \ref{fig:rpa}. In all cases a collective resonance appears in the RPA response, which is not trivial since we use a realistic NN-interaction. The centroid energies for the isoscalar giant monopole resonances are in nice agreement with experiment. Keeping in mind that there are no free parameters, this is a remarkable result indicating that the incompressibility generated by $\VO_{\UCOM}$ is reasonable. However, for the isovector dipole and the isoscalar quadrupole giant resonances the calculated centroid energies are systematically larger than the experimental ones. This hints at too small a value for the effective mass which is consistent with the very wide single-particle spectra resulting from the HF calculations \cite{RoPa05}. Again this could be an indication of a residual repulsive three-body force but also of the importance of additional correlations not included in the standard RPA framework.

\section{Conclusions \& Outlook}

We have used the correlated realistic interaction $\VO_{\UCOM}$ derived from the Argonne V18 potential as a universal starting point for nuclear structure calculations throughout the nuclear chart. Different many-body approaches ranging from no-core shell model to Hartree-Fock, many-body perturbation theory, and RPA have been employed using the same interaction. We observe that the binding energies per nucleon resulting from $\VO_{\UCOM}$ are in good agreement with experiment through the whole mass range, indicating that the impact of residual three-body forces on this observable is minimal. However, for other observables systematic deviations from experiment emerge minly in heavier nuclei: (\emph{i}) the rms-radii are too small, (\emph{ii}) the mean level-spacing of the HF single-particle spectra is too large leading to a too small effective mass, (\emph{iii}) the centroid energies of isovector dipole and isoscalar giant quadrupole resonances are overestimated. One possible origin of these deviations, besides missing long-range correlations, is a residual repulsive three-body interaction not included in the present calculations. Initial calculations using a simple phenomenological three-body contact interaction indicate that all aforementioned discrepancies can be reduced.  The inclusion of effective three-body interactions into the many-body schemes discussed here is the subject of ongoing investigations.


\end{document}